\documentclass[12pt,aps,prl,preprint,tightenlines,onecolumn,superscriptaddress,amsmath,amssymb,floatfix]{revtex4-1}
\usepackage{graphicx}
\def\eq{\begin{equation}}
\def\en{\end{equation}}
\def\aeq#1{\begin{align}#1\end{align}}  

\def\bdry{\mathit{bdry}}
\def\bulk{\mathit{bulk}}
\begin{document}
\preprint{IPMU12-0129}
\title{\vspace*{4ex}
Lower bound on the entropy of boundaries 
and junctions\\ in 1+1d quantum critical systems
\vspace*{3ex}}
\author{Daniel Friedan}
\affiliation{New High Energy Theory Center and Department of Physics and Astronomy,
Rutgers, The State University of New Jersey,
Piscataway, New Jersey 08854-8019 U.S.A.}
\affiliation{The Science Institute, The University of Iceland,
Reykjavik, Iceland}
\author{Anatoly Konechny}
\affiliation{Department of Mathematics, Heriot-Watt University,
EH14 4AS Edinburgh, United Kingdom}
\affiliation{Maxwell Institute for Mathematical Sciences, Edinburgh, United Kingdom}
\author{Cornelius Schmidt-Colinet
\vspace*{3ex}}
\affiliation{Kavli Institute for the Physics and Mathematics of the
Universe,
Todai Institutes for Advanced Study (TODIAS),
The University of Tokyo,
Kashiwa, Chiba 277-8582, Japan
\vspace*{6ex}}
%
%

\begin{abstract}
\vspace*{2ex}
A lower bound is derived for the boundary entropy $s=\ln g$ of a 
1+1d quantum critical system with boundary
under the conditions $c\ge 1$ on the bulk conformal central charge
and $\Delta_{1} > (c-1)/12$
on the most relevant bulk scaling dimension.
This is the first general restriction on the possible values of $g$ 
for bulk critical systems with $c\ge 1$.

\end{abstract}
\maketitle

A 1+1d quantum critical system is described by a 2d conformal field 
theory (the bulk CFT).
A critical boundary is described by a conformally invariant boundary 
condition on the bulk CFT.
The combination --- a bulk CFT with a conformally invariant boundary 
condition --- is a boundary CFT\cite{Cardy:1984bb}.
Critical junctions in critical quantum 
circuits are described by boundary CFT's.
For an $N$-wire junction,
the bulk CFT is the $N$-fold product of the CFT's describing the individual wires.
The
critical junction is described by a conformal boundary condition for the 
product CFT.
In string theory, branes in 
spacetime are described by conformal boundary conditions on the string 
worldsheet.

Affleck and Ludwig\cite{AL1} defined a number $g$ for each boundary CFT
--- the {\it universal non-integer ground state degeneracy}.
The entropy localized in the boundary ---
the boundary entropy --- is $\ln g$.
It is defined as
the total entropy of the system
minus the bulk entropy $\pi c L/6\beta$, which is proportional to the length 
$L$ in the limit of large $L$.
The coefficient of $L$  is determined by
conformal invariance, in terms of the inverse 
temperature $\beta$ and the conformal central charge $c$ of the bulk 
CFT.

For $c<1$, there is a complete classification
of all possible boundary CFT's\cite{Behrend:1999bn}.
There is also a complete classification of conformal boundary 
conditions for the $c=1$ gaussian 
model\cite{Friedan1999n,Gaberdiel:2001zq,Janik:2001hb}.
Until now, no limitations have been known on
the possible values that $g$ can take for any other $c\ge 1$ bulk 
systems.

For non-critical boundary conditions in a bulk CFT,
the boundary entropy $s$ is defined in the same way
by subtracting the universal bulk entropy from the total entropy.
Now $s$ depends on the temperature.  Under 
a change of the thermal length scale
$\beta$ the effective boundary condition evolves
along
the boundary renormalization group flow (the boundary RG flow).
The bulk system, being scale invariant, stays the same.  A 
fixed point of the boundary RG flow is a boundary CFT.
At a fixed point $s=\ln g$.  It is not obvious that $s$ decreases with 
decreasing temperature ---
that the second law of thermodynamics applies to the boundary ---
because of the subtraction 
of bulk entropy in the definition of $s$.
In fact, the boundary 
entropy $s$ does decrease along the boundary RG flow, so it decreases with 
temperature\cite{Friedan2004a}.
The result is actually stronger: the boundary RG beta function is the 
gradient of the function $s$ on the space of boundary conditions.  
All that is missing
to control the asymptotic 
low temperature behavior
is a lower bound on $s$.  Such a lower bound would be an
analogue of the third law of thermodynamics.
Again, the existence of a lower bound on $s$ is non-obvious because of
the subtraction of the bulk entropy.
Unsuccessful attempts have been made to 
prove that $s$ is bounded below\cite{Friedan2006a}.  
Without a lower bound, we cannot exclude the possibility that $s$ 
might decrease to $-\infty$ as the temperature drops to zero.

Here we prove a lower bound $g>g_{B}(c,\Delta_{1})$ 
that applies to any $c\ge 1$ bulk system that has $\Delta_{1} > (c-1)/12$
where $\Delta_{1}$ is the most relevant bulk scaling dimension.
The proof assumes nothing about the boundary condition besides 
criticality and unitarity.
The bound does not imply a
boundary third law of thermodynamics,
since it applies only to critical boundary conditions.
It does imply that
a non-critical boundary with entropy $s$ below 
the bound cannot flow to a critical boundary condition at zero 
temperature.  If such a system exists,
its boundary entropy must necessarily decrease 
without limit towards $s=-\infty$ at zero temperature.

One of us has argued that critical quantum circuits are
natural physical systems for asymptotically large scale quantum 
computers\cite{Friedan2005bc}.  The quantum wires should be critical in the bulk, so that
the low-energy excitations are protected against microscopic fluctuations by universality 
(the RG), and travel at uniform speed.
The processing elements are to be the 
circuit junctions.  
A junction can be considered as a boundary condition on the CFT describing the 
independent wires entering it.
A lower bound on $\ln g$ leads to an upper bound on the information 
capacity of the junction, giving
a general constraint on the design of critical 
quantum circuits.

In string theory, $g$ is the brane tension.  The lower bound 
on the brane tension might be useful
once it is extended 
to superconformal boundary CFT's
and if the condition $\Delta_{1}> (c-1)/12$ can be relaxed.



The modular duality formula for a boundary CFT is\cite{Cardy}
$$
\mathrm{tr}\, \exp{(-\beta H_{\bdry})} = 
\langle {B} | \, \exp({- 2\pi H_{\bulk}}/\beta)  \, |B \rangle.
$$
On the left is the thermodynamic
partition function $Z_{L}(\beta)$ at inverse temperature $\beta$
for a finite segment of the system of length $L=1$.
The boundary conditions at the two ends of the segment are the same.
The hamiltonian is $H_{\bdry}$.
The Hilbert space is called the 
\emph{boundary sector}
(in string theory, the \emph{open string sector}). 
In the euclidean space-time interpretation, $Z_{L}(\beta)$ 
is the partition function of a finite 2-d cylinder with length $L$ and 
euclidean time periodic with period $\beta$.
The right hand side is obtained by 
re-interpreting $L$ as euclidean time and $\beta$ as the length of a
circle or, by scale invariance, euclidean time $2\pi L/\beta$ and a 
spatial circle of length $2\pi$.
The hamiltonian for the circle is $H_{\bulk}$.
The Hilbert space of the bulk system on the circle is called the 
\emph{bulk sector} (the \emph{closed string sector}).
The boundary condition on each end of the cylinder
is described by a bulk state $|B\rangle$.
The modular duality formula states that
the partition function depends only on the 2-d geometry, so the two
quantum mechanical interpretations give the same result.

Conformal invariance implies that each side of the duality formula can be expressed 
as a sum over the characters of the irreducible unitary representations of the 
Virasoro algebra.  
For $c>1$ (we consider the case $c=1$ separately below)
the duality formula becomes
\eq
\chi_{0}(i\beta) + \sum_{j} \chi_{h_{j}}(i\beta)
= g^{2} \chi_{0}(i/\beta) + \sum_{k} b_{k}^{2} \,
\chi_{\Delta_{k}/2}(i/\beta)
\label{eq:duality1}
\nonumber
\en
where the characters $\chi_{h}(i\beta)$ are given by
$$
\chi_{h}(i\beta)= 
\frac{f_{h}(\beta)}{\eta(i\beta)},\quad
f_{h}(\beta)  = \left \{
\begin{array}{ll}
q^{-\gamma}(1-q), \quad & h=0
\\
q^{-\gamma+h},\quad & h>0,
\end{array}
\right .
$$
$$
q=e^{-2\pi\beta},\quad
\eta(i\beta) = q^{1/{24}}\prod_{n=1}^{\infty}(1-q^{n}),\quad
\gamma = \frac{c-1}{24}.
$$
The character $\chi_{0}(i\beta)$ is the contribution to the partition 
function from the
boundary sector representation that contains the ground state,
whose energy is $-2\pi c/24$.
The characters $\chi_{h_{j}}(i\beta)$ are the contributions from the representations with lowest 
energies $2\pi (h_{j}-c/24)$.  Unitarity and uniqueness of 
the ground state imply all $h_{j}>0$.
The boundary scaling fields are in one-to-one correspondence with
the energy eigenstates in the boundary sector, via radial quantization.
A \emph{primary} boundary field of scaling dimension $h_{j}$
corresponds to the lowest energy state in the representation labelled by $j$.
The bulk scaling fields are in 
one-to-one correspondence with 
the energy eigenstates in the bulk sector.
The terms on the right side of the duality formula come from the
closed sector representations whose lowest energy states
correspond to the spin-zero primary scaling fields
whose scaling dimensions are $0<\Delta_{1}\le \Delta_{2} \le \cdots$.
The numbers $g$ and $b_{k}$ characterize and
completely determine the conformally invariant boundary state 
$|B\rangle$.


Rattazzi, Rychkov, Tonni, and Vichi\cite{Rattazzi:2008pe}
developed the \emph{linear functional method} 
for deriving bounds on the 
low-lying scaling dimensions of 
conformal field theories from crossing formulas for correlation 
functions.  Hellerman\cite{Hellerman:2009bu} showed that the same method could be 
applied to the modular duality formula for the bulk partition 
function of a 2d CFT to obtain an upper bound on the dimension of the lowest non-trivial primary field,
and, with one of us, to obtain bounds on state 
degeneracies\cite{Hellerman:2010qd}.
Here we apply the linear functional method to the modular duality 
formula for boundary CFT to derive a lower bound on $g$.

We want a bound on $g$ that depends only on
properties of the bulk system
so it will apply to all possible 
critical boundary conditions for a given bulk critical system.
The derivation should
use only universal facts about the boundary condition: the uniqueness of 
the boundary sector
ground state and the positivity of the scaling dimensions $h_{j}$,
which follows from unitarity.
Otherwise, nothing should be assumed about the numbers $h_{j}$ or  $b_{k}$.

We start by multiplying both sides of the duality formula by
$\eta(i\beta)= \beta^{-1/2} \eta(i/\beta)$ to get
\eq
\label{eq:duality}
f_{0}+ \sum_{j}f_{h_{j}}
= g^{2}\tilde f_{0}+ \sum_{k}b_{k}^{2}\tilde f_{\Delta_{k}}
\en
where
$$
\tilde f_{\Delta} = \left \{
\begin{array}{ll}
\beta^{-1/2}\tilde q ^{-\gamma+\Delta/2}(1-\tilde q), \quad & \Delta =0
\\
\beta^{-1/2}\tilde q ^{-\gamma+\Delta/2},\quad & \Delta>0,
\end{array}
\right .
\quad \tilde q = e^{-{2\pi}/{\beta}}.
$$
Then we apply a linear functional --- a distribution $\rho(\beta)$ 
--- to both sides
of equation~(\ref{eq:duality}), giving
\eq
(\rho, f_{0}) +\sum_{j} (\rho,f_{h_{j}})
= g^{2} (\rho, \tilde f_{0}) + \sum_{k}  b_{k}^{2} \,(\rho, \tilde 
f_{\Delta_{k}})
\nonumber
\en
where
$
(\rho,F) = \int_{0}^{\infty}d\beta\,\rho(\beta)\, F(\beta)
\,.
$
If we can choose $\rho(\beta)$ so that
\aeq{
(\rho,f_{h})&\ge 0\,,\qquad 
\forall\,h>0
\label{eq:cond1}
\\
(\rho, \tilde f_{\Delta}) &\le 0 \,,\qquad 
\forall\, \Delta\ge\Delta_{1}
\label{eq:cond2}
}
then we get an inequality
\eq
g^{2}  \,  (\rho, \tilde f_{0}) \ge
(\rho, f_{0})
\,.
\en
Next, using the identity
\eq
\beta^{-1/2}\tilde q ^{-\gamma+\Delta/2} = \int_{-\infty}^{\infty}dy\, e^{-\pi \beta y^{2}+ 2\pi i y 
\sqrt{\Delta-2\gamma}} 
\label{eq:identity}
\en
we see that condition~(\ref{eq:cond1}) implies $(\rho, \tilde 
f_{0})>0$ so 
we have a lower bound on $g$,
\eq
g^{2} \ge g_{B}^{2}[\rho]= \frac{(\rho, f_{0})}{(\rho, \tilde f_{0})}
\,.
\en
Maximizing over all distributions $\rho(\beta)$ 
satisfying conditions~(\ref{eq:cond1}) and~(\ref{eq:cond2}),
we obtain the optimal bound
\eq
g^{2} \ge g_{B}^{2}(c,\Delta_{1})=\max\nolimits_{\rho} g_{B}^{2}[\rho]
\,.
\label{eq:bound}
\en
It is not obvious that there exists \emph{any} distribution $\rho(\beta)$ 
satisfying both conditions~(\ref{eq:cond1}) and~(\ref{eq:cond2}).
Using identity~(\ref{eq:identity}), condition~(\ref{eq:cond2}) 
requires
\eq
\int_{-\infty}^{\infty}dy\, (\rho, f_{\gamma+y^{2}/2})
 \,\cos \left ( 2\pi y \sqrt{\Delta_{1}-2\gamma}\right )
\le 0\,.
\label{greatformula!}
\en
If $\Delta_{1}\le 2\gamma$ 
this is incompatible with condition~(\ref{eq:cond1}).
So the linear functional method can give a bound only if $\Delta_{1}>
(c-1)/{12}$.

The next step is to approximate the space of distributions by 
distributions of the form
$
(\rho,F) = \mathcal{D}F(\beta)
$
where $\mathcal{D}$ is an $N^{\mathit{th}}$ order differential 
operator in $\beta$.
A bound $g_{B}^{2}(c,\Delta_{1},N,\beta)$ is obtained by taking
the maximum in equation~(\ref{eq:bound}) over
the differential operators of order $N$.
The bound can only improve as $N$ increases.
The partition function is real analytic in $\beta$ so we can expect
the limit $N\rightarrow \infty$ to exhaust the space of linear 
functionals for any choice of $\beta$,
giving the optimal bound $g_{B}^{2}(c,\Delta_{1})=
\lim_{N\rightarrow\infty}g_{B}^{2}(c,\Delta_{1},N,\beta)$.
We 
stop here at $N=1$,
contenting ourselves with finding any bound at all.
Elsewhere we will use the numerical techniques
of~\cite{Poland:2011ey} (semi-definite 
programming) to approximate the optimal bound from the linear 
functional method.

For $N=1$, we write the general first order operator
\eq
\mathcal{D} =  a_{0} +a_{1} \left ( 
-\frac1{2\pi}\frac{\partial}{\partial{\beta}} +\gamma\right )\,.
\nonumber
\en
Condition~(\ref{eq:cond1}) is  $a_{0},\, a_{1}\ge 0$.
There is no bound if $a_{1}=0$, and the bound does not change if we 
scale $\mathcal{D}$, so we might as well set $a_{1}=1$.
Condition~(\ref{eq:cond2}) then becomes
\eq
a_{0} \le A_{1}(\beta) = 
\frac{\Delta_{1}-2\gamma}{2\beta^{2}} -\frac1{4\pi\beta}-\gamma
\,.
\nonumber
\en
These conditions require $A_{1}(\beta)\ge 0$ which cannot be
satisfied for any value of $\beta$ if $\Delta_{1}- 2\gamma\le 0$, so
to get a bound we have to assume $\Delta_{1}>2\gamma$, the necessity 
of which we have
already seen from the general analysis.
The bound~(\ref{eq:bound}) is
\eq
g_{B}^{2}[\rho] = A_{2}(\beta)
\frac{a_{0}-A_{3}(\beta)}{a_{0}+A_{4}(\beta)}
\en
where
\eq
A_{2}(\beta) = \beta^{\frac12} q^{-\gamma}\tilde q^{\gamma}
\frac{1-q}{1-\tilde q}
\,,\qquad
A_{3}(\beta) = \frac{q}{1-q},
\nonumber
\en
\eq
A_{4}(\beta) = \gamma +\frac1{4\pi\beta} + \frac{\gamma}{\beta^{2}}
+ \frac{1}{\beta^{2}} \frac{\tilde q}{1-\tilde q}
\,.
\nonumber
\en
Since $A_{2,3,4}(\beta) > 0$,
the highest bound is obtained when $a_{0}$ takes its maximum value
$A_{1}(\beta)$, so
\eq
g_{B}^{2}(c,\Delta_{1},1,\beta)  = A_{2}(\beta)
\frac{A_{1}(\beta)-A_{3}(\beta)}{A_{1}(\beta)+A_{4}(\beta)}
\,.
\en
The bound is empty unless $A_{1}(\beta)-A_{3}(\beta)>0$, which is stronger than 
$A_{1}(\beta)\ge 0$, so
\eq
A_{1}(\beta)-A_{3}(\beta)>0
\label{eq:cond3}
\en
is the only condition we need to impose to get a
bound.

At this point there is no reason to stick to one particular value of 
$\beta$.  The dependence on $\beta$ will 
disappear as $N\rightarrow \infty$ but for finite $N$ we can sample 
a larger subspace of distributions if we vary $\beta$.  The best 
bound that can be obtained with a first-order $\mathcal{D}$ is
\eq
g_{B}^{2}(c,\Delta_{1},1) = \max_{\beta} \,g_{B}^{2}(c,\Delta_{1},1,\beta)
\en
where the maximum is taken over all $\beta$ satisfying condition~(\ref{eq:cond3}).
There is a unique positive solution $\beta_{1}$ of 
$
A_{1}(\beta_{1})-A_{3}(\beta_{1})=0
$
and condition~(\ref{eq:cond3}) is equivalent to 
$0<\beta<\beta_{1}$.
So for $\Delta_{1}>2\gamma$ there is a lower bound
\eq
g^{2} \ge g_{B}^{2}(c,\Delta_{1},1)
\nonumber
\en
with
\eq
g_{B}^{2}(c,\Delta_{1},1) = \max_{0<\beta<\beta_{1}} A_{2}(\beta)
\frac{A_{1}(\beta)-A_{3}(\beta)}{A_{1}(\beta)+A_{4}(\beta)}\,.
\en
There is no analytic expression for the $N=1$ bound, but it can be 
calculated numerically for any given value of $c$ and $\Delta_{1}$.
In general,
the detailed form of the $N=1$ bound as a function of $c$ and $\Delta_{1}$ 
is not particularly interesting since it is not even the optimal 
linear functional bound.
At this stage, 
we are only interested in the fact that there is any lower bound on $g$.


\begin{figure}[h]
\begin{center}
\includegraphics[width=8.6cm]{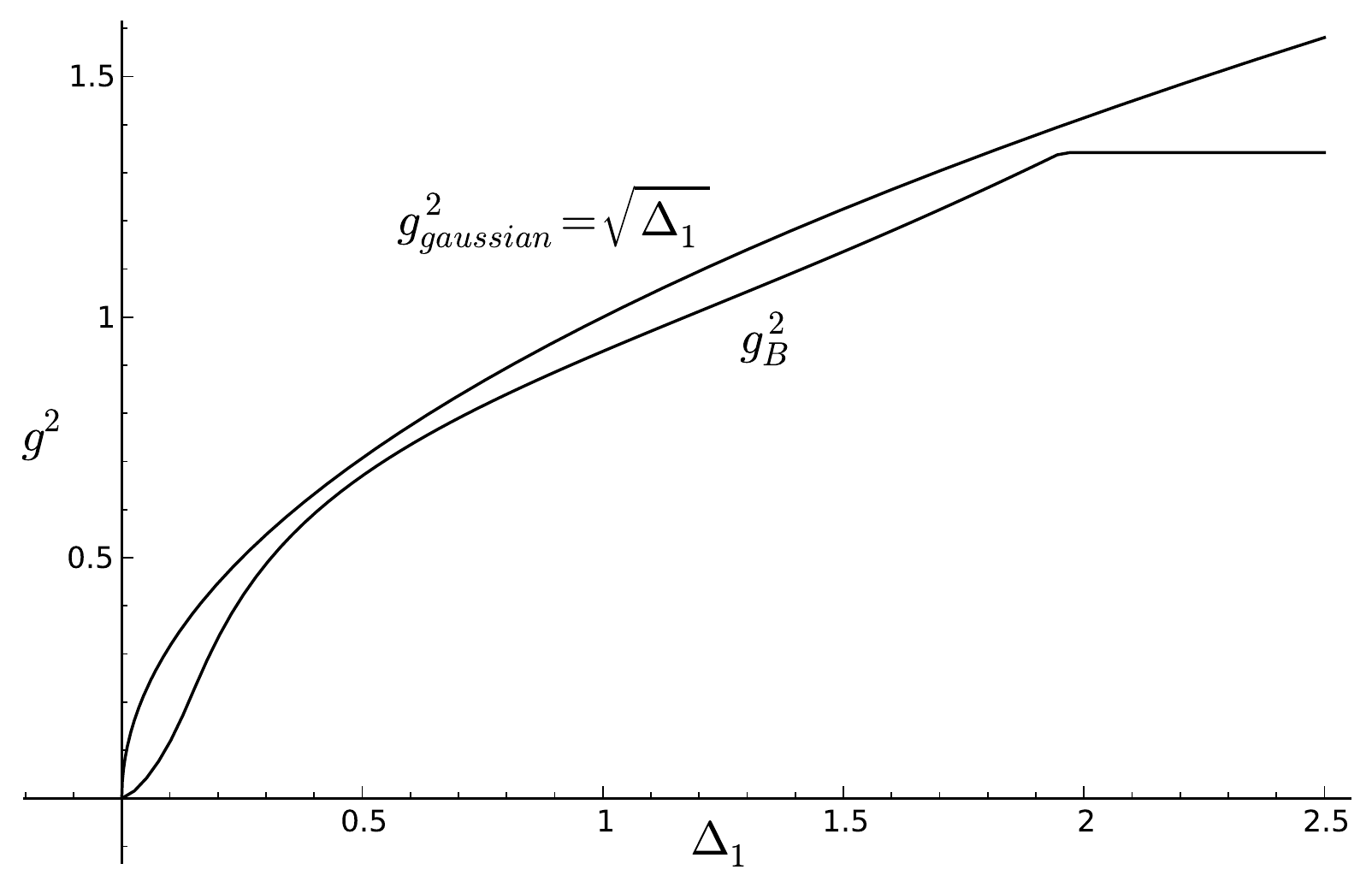}
\caption{\label{fig:c=1}
The $N=1$ bound for $c=1$ compared to the minimum value of
$g^{2}$ for the $c=1$ gaussian model. 
The comparison is extended past the maximal value $\Delta_1=1/2$ by
interpreting $\Delta_1$ as the lowest dimension of a primary occurring in the boundary state.}
\end{center}
\end{figure}
For $c=1$, there are degenerate Virasoro representations that do not 
occur for $c>1$.  
Equation~(\ref{eq:duality}) holds with the modification that,
for integers $n\ge 1$,
$f_{n^{2}} = q^{-\gamma+n^{2}}(1-q^{2n+1})$, 
$\tilde f_{2 n^{2}} = \beta^{-1/2}\tilde q^{-\gamma+n^{2}}(1-\tilde 
q^{2n+1})$.
As before, we apply a first order differential operator with 
appropriate positivity conditions to get a lower bound on $g$ that 
depends on $\beta$, then we maximize over $\beta$.
We omit the calculations.  The result is shown in
Figure~\ref{fig:c=1}.
Included for comparison is the smallest value of $g^{2}$ for the 
$c=1$ gaussian model.
The $N=1$ bound is moderately good except when $\Delta_{1}\approx 0$.


Several future directions are more or less obvious.
We can explore how much the bound can be improved by numerically 
maximizing over differential operators of degree $N>1$.
We can apply the linear functional method to supersymmetric CFT's to get bounds on brane 
tensions in superstring theory.
We can try to find linear functional bounds for specific bulk CFT's 
by exploiting knowledge of the bulk spectrum.
For example, the most interesting bulk universality class for critical quantum 
circuits is the Monster CFT\cite{Frenkel_MR747596},
which has $c=24$ and $\Delta_{1}=4$.
It is interesting because it has no relevant or marginal 
bulk perturbations.
Our $N=1$ lower bound is $g_{B}^{2}(24,4,1) = 0.0273+$.
The known conformal boundary conditions\cite{Craps:2002rw} have 
$g^{2}=1$.
Preliminary results of numerical calculations for $N$ up to 37, using the fact 
that all the bulk scaling dimensions $\Delta_{k}$ are integers 
$\ge 4$, give a bound $g^{2}>0.96$ which is strikingly close to $1$.

The most pressing problem is to overcome the restriction $\Delta_{1}> 
(c-1)/{12}$.
We expect ---
from consideration of the effective low energy field 
theory of string theory in the presence of 
branes 
---
that there should be a lower bound on $g$ for all $\Delta_{1}$ 
which goes to zero as $\Delta_{1}$ goes to zero.
We have shown that in consequence of \eqref{greatformula!}
our present method cannot be extended straightforwardly.
Some new ideas will be needed.
The linear functional method applied to the boundary partition function
is a practical compromise, well short of the exact lower bound that 
would follow from a complete solution of the conformal bootstrap for 
boundary CFT.
We do not know in what direction to improve the linear functional 
method to get past the restriction $\Delta_{1}> 
(c-1)/{12}$.

C.S.C thanks S. Hellerman and D.F. thanks A. Vichi and E. Diaconescu for discussions.
D.F. and C.S.C. would like to thank Heriot-Watt University for hospitality 
during the initial stages of this project.
The work of D.F. was supported by the Rutgers New High Energy Theory Center
and by U.S. Department of Energy grant DE-FG02-12ER41813. 
A.K. acknowledges the support of the STFC grants ST/G000514/1 ``String Theory Scotland''
and ST/J000310/1 ``High energy physics at the Tait Institute''.
C.S.C. was supported by the WPI Research Center Initiative, MEXT, Japan.
\bibliography{Literature}

\begin{thebibliography}{16}%
\makeatletter
\providecommand \@ifxundefined [1]{%
 \@ifx{#1\undefined}
}%
\providecommand \@ifnum [1]{%
 \ifnum #1\expandafter \@firstoftwo
 \else \expandafter \@secondoftwo
 \fi
}%
\providecommand \@ifx [1]{%
 \ifx #1\expandafter \@firstoftwo
 \else \expandafter \@secondoftwo
 \fi
}%
\providecommand \natexlab [1]{#1}%
\providecommand \enquote  [1]{``#1''}%
\providecommand \bibnamefont  [1]{#1}%
\providecommand \bibfnamefont [1]{#1}%
\providecommand \citenamefont [1]{#1}%
\providecommand \href@noop [0]{\@secondoftwo}%
\providecommand \href [0]{\begingroup \@sanitize@url \@href}%
\providecommand \@href[1]{\@@startlink{#1}\@@href}%
\providecommand \@@href[1]{\endgroup#1\@@endlink}%
\providecommand \@sanitize@url [0]{\catcode `\\12\catcode `\$12\catcode
  `\&12\catcode `\#12\catcode `\^12\catcode `\_12\catcode `\%12\relax}%
\providecommand \@@startlink[1]{}%
\providecommand \@@endlink[0]{}%
\providecommand \url  [0]{\begingroup\@sanitize@url \@url }%
\providecommand \@url [1]{\endgroup\@href {#1}{\urlprefix }}%
\providecommand \urlprefix  [0]{URL }%
\providecommand \Eprint [0]{\href }%
\providecommand \doibase [0]{http://dx.doi.org/}%
\providecommand \selectlanguage [0]{\@gobble}%
\providecommand \bibinfo  [0]{\@secondoftwo}%
\providecommand \bibfield  [0]{\@secondoftwo}%
\providecommand \translation [1]{[#1]}%
\providecommand \BibitemOpen [0]{}%
\providecommand \bibitemStop [0]{}%
\providecommand \bibitemNoStop [0]{.\EOS\space}%
\providecommand \EOS [0]{\spacefactor3000\relax}%
\providecommand \BibitemShut  [1]{\csname bibitem#1\endcsname}%
\let\auto@bib@innerbib\@empty
\bibitem [{\citenamefont {Cardy}(1984)}]{Cardy:1984bb}%
  \BibitemOpen
  \bibfield  {author} {\bibinfo {author} {\bibfnamefont {J.~L.}\ \bibnamefont
  {Cardy}},\ }\href {\doibase 10.1016/0550-3213(84)90241-4} {\bibfield
  {journal} {\bibinfo  {journal} {Nucl.Phys.}\ }\textbf {\bibinfo {volume}
  {B240}},\ \bibinfo {pages} {514} (\bibinfo {year} {1984})}\BibitemShut
  {NoStop}%
\bibitem [{\citenamefont {Affleck}\ and\ \citenamefont {Ludwig}(1991)}]{AL1}%
  \BibitemOpen
  \bibfield  {author} {\bibinfo {author} {\bibfnamefont {I.}~\bibnamefont
  {Affleck}}\ and\ \bibinfo {author} {\bibfnamefont {A.~W.~W.}\ \bibnamefont
  {Ludwig}},\ }\href@noop {} {\bibfield  {journal} {\bibinfo  {journal} {Phys.
  Rev. Lett.}\ }\textbf {\bibinfo {volume} {67}},\ \bibinfo {pages} {161}
  (\bibinfo {year} {1991})}\BibitemShut {NoStop}%
\bibitem [{\citenamefont {Behrend}\ \emph {et~al.}(2000)\citenamefont
  {Behrend}, \citenamefont {Pearce}, \citenamefont {Petkova},\ and\
  \citenamefont {Zuber}}]{Behrend:1999bn}%
  \BibitemOpen
  \bibfield  {author} {\bibinfo {author} {\bibfnamefont {R.~E.}\ \bibnamefont
  {Behrend}}, \bibinfo {author} {\bibfnamefont {P.~A.}\ \bibnamefont {Pearce}},
  \bibinfo {author} {\bibfnamefont {V.~B.}\ \bibnamefont {Petkova}}, \ and\
  \bibinfo {author} {\bibfnamefont {J.-B.}\ \bibnamefont {Zuber}},\ }\href
  {\doibase 10.1016/S0550-3213(99)00592-1, 10.1016/S0550-3213(99)00592-1}
  {\bibfield  {journal} {\bibinfo  {journal} {Nucl.Phys.}\ }\textbf {\bibinfo
  {volume} {B570}},\ \bibinfo {pages} {525} (\bibinfo {year} {2000})},\ \Eprint
  {http://arxiv.org/abs/hep-th/9908036} {arXiv:hep-th/9908036} \BibitemShut
  {NoStop}%
\bibitem [{\citenamefont {Friedan}(1999)}]{Friedan1999n}%
  \BibitemOpen
  \bibfield  {author} {\bibinfo {author} {\bibfnamefont {D.}~\bibnamefont
  {Friedan}},\ }\href
  {http://www.physics.rutgers.edu/pages/friedan/papers/boundary_c=1_1999.pdf}
  {\enquote {\bibinfo {title} {The space of conformal boundary conditions for
  the c=1 gaussian model}}} (\bibinfo {year} {1999}),\ \bibinfo {note}
  {{unpublished, http://www.physics.rutgers.edu/pages/friedan/}}\BibitemShut
  {NoStop}%
\bibitem [{\citenamefont {Gaberdiel}\ and\ \citenamefont
  {Recknagel}(2001)}]{Gaberdiel:2001zq}%
  \BibitemOpen
  \bibfield  {author} {\bibinfo {author} {\bibfnamefont {M.}~\bibnamefont
  {Gaberdiel}}\ and\ \bibinfo {author} {\bibfnamefont {A.}~\bibnamefont
  {Recknagel}},\ }\href@noop {} {\bibfield  {journal} {\bibinfo  {journal}
  {JHEP}\ }\textbf {\bibinfo {volume} {0111}},\ \bibinfo {pages} {016}
  (\bibinfo {year} {2001})},\ \Eprint {http://arxiv.org/abs/hep-th/0108238}
  {arXiv:hep-th/0108238} \BibitemShut {NoStop}%
\bibitem [{\citenamefont {Janik}(2001)}]{Janik:2001hb}%
  \BibitemOpen
  \bibfield  {author} {\bibinfo {author} {\bibfnamefont {R.~A.}\ \bibnamefont
  {Janik}},\ }\href {\doibase 10.1016/S0550-3213(01)00486-2} {\bibfield
  {journal} {\bibinfo  {journal} {Nucl.Phys.}\ }\textbf {\bibinfo {volume}
  {B618}},\ \bibinfo {pages} {675} (\bibinfo {year} {2001})},\ \Eprint
  {http://arxiv.org/abs/hep-th/0109021} {arXiv:hep-th/0109021} \BibitemShut
  {NoStop}%
\bibitem [{\citenamefont {Friedan}\ and\ \citenamefont
  {Konechny}(2004)}]{Friedan2004a}%
  \BibitemOpen
  \bibfield  {author} {\bibinfo {author} {\bibfnamefont {D.}~\bibnamefont
  {Friedan}}\ and\ \bibinfo {author} {\bibfnamefont {A.}~\bibnamefont
  {Konechny}},\ }\href {\doibase 10.1103/PhysRevLett.93.030402} {\bibfield
  {journal} {\bibinfo  {journal} {Phys. Rev. Lett.}\ }\textbf {\bibinfo
  {volume} {93}},\ \bibinfo {pages} {030402} (\bibinfo {year} {2004})},\
  \Eprint {http://arxiv.org/abs/hep-th/0312197} {arXiv:hep-th/0312197}
  \BibitemShut {NoStop}%
\bibitem [{\citenamefont {Friedan}\ and\ \citenamefont
  {Konechny}(2006)}]{Friedan2006a}%
  \BibitemOpen
  \bibfield  {author} {\bibinfo {author} {\bibfnamefont {D.}~\bibnamefont
  {Friedan}}\ and\ \bibinfo {author} {\bibfnamefont {A.}~\bibnamefont
  {Konechny}},\ }\href@noop {} {\bibfield  {journal} {\bibinfo  {journal} {J.
  Stat. Phys.}\ }\textbf {\bibinfo {volume} {0603}},\ \bibinfo {pages} {P014}
  (\bibinfo {year} {2006})},\ \Eprint {http://arxiv.org/abs/hep-th/0512023}
  {arXiv:hep-th/0512023} \BibitemShut {NoStop}%
\bibitem [{\citenamefont {Friedan}(2005)}]{Friedan2005bc}%
  \BibitemOpen
  \bibfield  {author} {\bibinfo {author} {\bibfnamefont {D.}~\bibnamefont
  {Friedan}},\ }\href@noop {} {\bibfield  {journal} {\bibinfo  {journal}
  {{{arXiv:cond-mat/0505084, 05050845}}}} (\bibinfo {year}
  {2005})}\BibitemShut {NoStop}%
\bibitem [{\citenamefont {Cardy}(1989)}]{Cardy}%
  \BibitemOpen
  \bibfield  {author} {\bibinfo {author} {\bibfnamefont {J.~L.}\ \bibnamefont
  {Cardy}},\ }\href@noop {} {\bibfield  {journal} {\bibinfo  {journal} {Nucl.
  Phys.}\ }\textbf {\bibinfo {volume} {B324}},\ \bibinfo {pages} {581}
  (\bibinfo {year} {1989})}\BibitemShut {NoStop}%
\bibitem [{\citenamefont {Rattazzi}\ \emph {et~al.}(2008)\citenamefont
  {Rattazzi}, \citenamefont {Rychkov}, \citenamefont {Tonni},\ and\
  \citenamefont {Vichi}}]{Rattazzi:2008pe}%
  \BibitemOpen
  \bibfield  {author} {\bibinfo {author} {\bibfnamefont {R.}~\bibnamefont
  {Rattazzi}}, \bibinfo {author} {\bibfnamefont {V.~S.}\ \bibnamefont
  {Rychkov}}, \bibinfo {author} {\bibfnamefont {E.}~\bibnamefont {Tonni}}, \
  and\ \bibinfo {author} {\bibfnamefont {A.}~\bibnamefont {Vichi}},\ }\href
  {\doibase 10.1088/1126-6708/2008/12/031} {\bibfield  {journal} {\bibinfo
  {journal} {JHEP}\ }\textbf {\bibinfo {volume} {0812}},\ \bibinfo {pages}
  {031} (\bibinfo {year} {2008})},\ \Eprint {http://arxiv.org/abs/0807.0004}
  {arXiv:0807.0004 [hep-th]} \BibitemShut {NoStop}%
\bibitem [{\citenamefont {Hellerman}(2011)}]{Hellerman:2009bu}%
  \BibitemOpen
  \bibfield  {author} {\bibinfo {author} {\bibfnamefont {S.}~\bibnamefont
  {Hellerman}},\ }\href {\doibase 10.1007/JHEP08(2011)130} {\bibfield
  {journal} {\bibinfo  {journal} {JHEP}\ }\textbf {\bibinfo {volume} {1108}},\
  \bibinfo {pages} {130} (\bibinfo {year} {2011})},\ \Eprint
  {http://arxiv.org/abs/0902.2790} {arXiv:0902.2790 [hep-th]} \BibitemShut
  {NoStop}%
\bibitem [{\citenamefont {Hellerman}\ and\ \citenamefont
  {Schmidt-Colinet}(2011)}]{Hellerman:2010qd}%
  \BibitemOpen
  \bibfield  {author} {\bibinfo {author} {\bibfnamefont {S.}~\bibnamefont
  {Hellerman}}\ and\ \bibinfo {author} {\bibfnamefont {C.}~\bibnamefont
  {Schmidt-Colinet}},\ }\href {\doibase 10.1007/JHEP08(2011)127} {\bibfield
  {journal} {\bibinfo  {journal} {JHEP}\ }\textbf {\bibinfo {volume} {1108}},\
  \bibinfo {pages} {127} (\bibinfo {year} {2011})},\ \Eprint
  {http://arxiv.org/abs/1007.0756} {arXiv:1007.0756 [hep-th]} \BibitemShut
  {NoStop}%
\bibitem [{\citenamefont {Poland}\ \emph {et~al.}(2012)\citenamefont {Poland},
  \citenamefont {Simmons-Duffin},\ and\ \citenamefont {Vichi}}]{Poland:2011ey}%
  \BibitemOpen
  \bibfield  {author} {\bibinfo {author} {\bibfnamefont {D.}~\bibnamefont
  {Poland}}, \bibinfo {author} {\bibfnamefont {D.}~\bibnamefont
  {Simmons-Duffin}}, \ and\ \bibinfo {author} {\bibfnamefont {A.}~\bibnamefont
  {Vichi}},\ }\href@noop {} {\bibfield  {journal} {\bibinfo  {journal} {JHEP}\
  }\textbf {\bibinfo {volume} {1205}},\ \bibinfo {pages} {110} (\bibinfo {year}
  {2012})},\ \Eprint {http://arxiv.org/abs/1109.5176} {arXiv:1109.5176
  [hep-th]} \BibitemShut {NoStop}%
\bibitem [{\citenamefont {Frenkel}\ \emph {et~al.}(1984)\citenamefont
  {Frenkel}, \citenamefont {Lepowsky},\ and\ \citenamefont
  {Meurman}}]{Frenkel_MR747596}%
  \BibitemOpen
  \bibfield  {author} {\bibinfo {author} {\bibfnamefont {I.~B.}\ \bibnamefont
  {Frenkel}}, \bibinfo {author} {\bibfnamefont {J.}~\bibnamefont {Lepowsky}}, \
  and\ \bibinfo {author} {\bibfnamefont {A.}~\bibnamefont {Meurman}},\ }\href
  {\doibase 10.1073/pnas.81.10.3256} {\bibfield  {journal} {\bibinfo  {journal}
  {Proc. Nat. Acad. Sci. U.S.A.}\ }\textbf {\bibinfo {volume} {81}},\ \bibinfo
  {pages} {3256} (\bibinfo {year} {1984})}\BibitemShut {NoStop}%
\bibitem [{\citenamefont {Craps}\ \emph {et~al.}(2003)\citenamefont {Craps},
  \citenamefont {Gaberdiel},\ and\ \citenamefont {Harvey}}]{Craps:2002rw}%
  \BibitemOpen
  \bibfield  {author} {\bibinfo {author} {\bibfnamefont {B.}~\bibnamefont
  {Craps}}, \bibinfo {author} {\bibfnamefont {M.~R.}\ \bibnamefont
  {Gaberdiel}}, \ and\ \bibinfo {author} {\bibfnamefont {J.~A.}\ \bibnamefont
  {Harvey}},\ }\href {\doibase 10.1007/s00220-002-0763-7} {\bibfield  {journal}
  {\bibinfo  {journal} {Commun.Math.Phys.}\ }\textbf {\bibinfo {volume}
  {234}},\ \bibinfo {pages} {229} (\bibinfo {year} {2003})},\ \Eprint
  {http://arxiv.org/abs/hep-th/0202074} {arXiv:hep-th/0202074} \BibitemShut
  {NoStop}%
\end{thebibliography}%

\end{document}